\documentclass[twocolumn]{aastex631}

\usepackage{lipsum}
\usepackage{comment}
\usepackage{graphicx}	
\usepackage{amsmath}	
\usepackage{amssymb}	
\usepackage{xspace}
\usepackage{subfigure}
\usepackage{relsize} 
\usepackage{longtable}
\usepackage{xcolor}

\usepackage{array,etoolbox}
\preto\tabular{\setcounter{magicrownumbers}{0}}
\newcounter{magicrownumbers}

\usepackage{ragged2e}

\shorttitle{}
\shortauthors{Foord et al.}
\usepackage[mathcal]{eucal}

\newcommand{\BAYMAX}{\texttt{BAYMAX}\xspace}
\newcommand{\BF}{\mathcal{BF}\xspace}
\newcommand{\logBF}{\log{\BF}\xspace}

\usepackage{floatrow}
\usepackage{natbib}
\usepackage{booktabs}
\bibpunct{(}{)}{;}{a}{}{,}
\def\bibsection{}
\usepackage{paralist}
\usepackage{sidecap,wrapfig}
\usepackage[pdftex]{thumbpdf}
\usepackage{relsize}
\usepackage{mathptmx}
\usepackage{enumitem}
\usepackage{epstopdf}
\DeclareGraphicsExtensions{.pdf,.png,.jpg,.jpeg,.gif}
\providecommand{\ion}[2]{#1$\;$\textsmaller{\@Roman{#2}}}

\begin{document}

\title{Chandra Discovery of a Candidate Hyper-Luminous X-ray Source in  MCG+11-11-032}

\author[0000-0002-1616-1701]{Adi Foord}
\affil{Department of Physics, University of Maryland Baltimore County, 1000 Hilltop Cir, Baltimore, MD 21250, USA}

\author[0000-0002-2115-1137]{Francesca Civano}
\affil{NASA Goddard Space Flight Center, Greenbelt, MD 20771, USA}

\author[0000-0001-8627-4907]{Julia M.~Comerford}
\affil{Department of Astrophysical and Planetary Sciences, University of Colorado, Boulder, CO 80309, USA}

\author[0000-0001-5060-1398]{Martin Elvis}
\affil{Harvard-Smithsonian Center for Astrophysics, 60 Garden St., Cambridge, MA 02138, USA}

\author[0000-0002-3554-3318]{Giuseppina Fabbiano}
\affil{Harvard-Smithsonian Center for Astrophysics, 60 Garden St., Cambridge, MA 02138, USA}

\author[0000-0001-5766-4287]{Tingting Liu}
\affiliation{Department of Physics and Astronomy, West Virginia University, P.O. Box 6315, Morgantown, WV 26506, USA}

\author[0000-0003-0083-1157]{Elisabeta Lusso}
\affil{Dipartamento di Fisica e Astronomia, Universit\'{a} di Firenze, Via G. Sansone 1, 50019 Sesto Fiorentino, FI, Italy}
\affil{INAF - Osservatorio Astrofisico di Arcetri, Largo E.Fermi 5, 50125 Firenze, Italy}

\author[0000-0001-5544-0749]{Stefano Marchesi}
\affiliation{Dipartimento di Fisica e Astronomia (DIFA), Università di Bologna, via Gobetti 93/2, I-40129 Bologna, Italy}
\affiliation{Department of Physics and Astronomy, Clemson University, Kinard Lab of Physics, Clemson, SC 29634-0978, USA}
\affiliation{INAF - Osservatorio di Astrofisica e Scienza dello Spazio di Bologna, Via Piero Gobetti, 93/3, 40129, Bologna, Italy}

\author[0000-0003-4440-259X]{Mar Mezcua}
\affiliation{Institute of Space Sciences (ICE, CSIC), Campus UAB, Carrer de Magrans, 08193 Barcelona, Spain}
\affiliation{Institut d'Estudis Espacials de Catalunya (IEEC), Edifici RDIT, Campus UPC, 08860 Castelldefels (Barcelona), Spain}

\author[0000-0002-2713-0628]{Francisco Muller-Sanchez}
\affiliation{Department of Physics and Materials Science, The University of Memphis, 3720 Alumni Avenue, Memphis, TN 38152, USA}

\author[0000-0003-1056-8401]{Rebecca Nevin}
\affil{Fermi National Accelerator Laboratory, P.O. Box 500, Batavia, IL 60510, USA}

\author[0000-0003-1991-370X]{Kristina Nyland}
\affiliation{U.S. Naval Research Laboratory, 4555 Overlook Ave. SW, Washington, DC 20375, USA}



\begin{abstract}
\noindent We present a multi-wavelength analysis of MCG+11-11-032, a nearby AGN with the unique classification of both a binary and a dual AGN candidate. With new Chandra observations we aim to resolve any dual AGN system via imaging data, and search for signs of a binary AGN via analysis of the X-ray spectrum. Analyzing the Chandra spectrum, we find no evidence of previously suggested double-peaked Fe K$\alpha$ lines; the spectrum is instead best fit by an absorbed powerlaw with a single Fe K$\alpha$ line, as well as an additional line centered at $\approx$7.5 keV. The Chandra observation reveals faint, soft, and extended X-ray emission, possibly linked to low-level nuclear outflows. Further analysis shows evidence for a compact, hard source -- MCG+11-11-032 X2 -- located 3.27\arcsec\ from the primary AGN. Modeling MCG+11-11-032 X2 as a compact source, we find that it is relatively luminous ($L_{\text{2$-$10 keV}} = 1.52_{-0.48}^{+0.96}\times 10^{41}$ erg s$^{-1}$), and the location is coincident with an compact and off-nuclear source resolved in Hubble Space Telescope infrared (F105W) and ultraviolet (F621M, F547M) bands. Pairing our X-ray results with a 144 MHz radio detection at the host galaxy location, we observe X-ray and radio properties similar to those of ESO 243-49 HLX-1, suggesting that MCG+11-11-032 X2 may be a hyper-luminous X-ray source. This detection with Chandra highlights the importance of a high-resolution X-ray imager, and how previous binary AGN candidates detected with large-aperture instruments benefit from high-resolution follow-up. Future spatially resolved optical spectra, and deeper X-ray observations, can better constrain the origin of MCG+11-11-032 X2.
\end{abstract}



\section{Introduction} \label{sec:intro}
MCG+11-11-032 is a nearby ($z=0.0362$) Seyfert 2 galaxy, where the central active galactic nucleus (AGN) has the unique classification of both a ``dual AGN'' candidate \citep{Comerford2012} and a ``binary AGN'' candidate \citep{Severgnini2018}. Dual and binary AGN are a natural consequence of galaxy-galaxy mergers (e.g. \citealt{Volonteri2008, HopkinsMergerTriggerAGN, ColpiandDotti2011}). During stages of a galaxy merger where both central supermassive black holes (SMBHs) are brightly accreting as AGN, the system can be classified as a dual AGN during the earliest stages (at kiloparsec-scale separations), and a binary AGN at the final stages (where it has evolved down to sub-parsec separations; \citealt{Begelman2002}).

The dual AGN classification stems from double-peaked [\ion{O}{3}] emission lines discovered in the Sloan Digital Sky Survey (SDSS) spectrum, with shifts in central wavelength that are inconsistent with the host galaxy redshift \citep{Wang2009}. The source was then analyzed in \cite{Comerford2012} using the Blue Channel Spectrograph
on the MMT 6.5 m telescope. \cite{Comerford2012} report a velocity separation between the two [\ion{O}{3}] peaks of 275 km s$^{-1}$ and a projected separation on the sky of 0.77\arcsec\ (corresponding to 0.55 kpc at $z=0.0362$). The two line peaks were reported to be maximally separated at a position angle close to that of the galaxy's plane (PA$_{\text{opt}}=46^{\circ}$).~\cite{Severgnini2018} performed a second analysis of the SDSS spectrum of MCG+11-11-032, confirming the results found by \cite{Comerford2012} and reaching similar conclusions for H$\alpha$, [\ion{N}{2}] and [\ion{S}{2}] emission line velocity offsets. These results suggest the possibility of two distinct narrow line regions associated with a dual AGN, although other possibilities include a single AGN with outflows, disk rotation, and/or dust obscuration (e.g., \citealt{Nevin2016}).

\cite{Severgnini2018} analyzed available Swift X-Ray Telescope (XRT) and Burst Alert Telescope (BAT) observations to further investigate the source.  A variability analysis using Swift XRT (54 days of observations over a period of approximately 1 year) shows the source varying between a bright and faint state (XRT count rates: bright, 0.03 cts/s; faint, 0.01 cts/s). Because the spectral properties remained consistent between the two states, \cite{Severgnini2018} concluded that the observed variability may be due to intrinsic flux variations, versus variable levels of absorption along the line-of-sight. The best-fit model to the XRT X-ray spectrum hinted at the presence of two emission lines at $\sim$6.16 keV ($3\sigma$) and $\sim$ 6.56 keV ($<$ 2$\sigma$), respectively. An analysis of the 123-month 15$-$150 keV BAT light curve showed evidence for a sinusoidal signal, with a period of approximately 25 months. 

The velocity separation of the two emission lines (0.4 keV or $\sim0.06c$) is inconsistent with the projected separation measured from the double-peaked optical narrow emission lines. \cite{Severgnini2018} proposed an alternative interpretation for this source, where the $\sim$25-month periodicity of the X-ray emission may be due to a sub-parsec binary SMBH system. Many numerical simulations have been performed to understand the nature of binary SMBH variability, finding that sub-parsec binary SMBHs can form a circumbinary disk (e.g., \citealt{MacFadyen2008, DOrazio2013, Farris2015, DOrzaio2016, Tang2018, Moody2019, Tiede2020, Duffell2020, DOrazio2021, Derdzinski2021, Whitley2024}). The mass accretion rate of this disk is modulated by the orbital period of the SMBH binary, resulting in observed periodic X-ray emission (e.g., \citealt{DOrazio2013, Gold2014, Farris2014}). Assuming a 25-month period and a mass of 5$\times$10$^8$ M$_{\odot}$ (as measured from the velocity dispersion of the CO emission line; \citealt{Lamperti2017}), \citealt{Severgnini2018} calculate an orbital velocity of $\sim0.06c$, consistent with the velocity difference measured using the Fe K$\alpha$ emission line profiles in the X-ray spectrum. \cite{Liu2020} present an analysis using publically available BAT light curve data of MCG+11-11-032 from the Third Palermo BAT Catalog (3PBC). They find that a sinusoidal signal is rejected at $>90\%$ confidence interval; however, given that 3PBC only includes the first 66 months of the survey, robustly detecting two cycles of the signal is more difficult.

MCG+11-11-032 is thus a peculiar source showing evidence for double-peaked emission lines in the optical and (tentatively) the X-ray spectra. Confirmation of either a dual or binary AGN nature can be achieved with the Chandra X-ray Observatory. If the observed narrow emission lines presented in \cite{Comerford2012} are possibly produced by two AGN separated by 0.77\arcsec, then Chandra can resolve them as two distinct X-ray sources. Alternatively, follow-up Chandra observations of the spectrum of MCG+11-11-032 allow for the confirmation of two narrow Fe emission lines near $\sim$ 6.4 keV as presented in \citep{Severgnini2018}. The sub-arcsecond spatial resolution and energy resolution of Chandra are necessary to better constrain the X-ray properties of MCG+11-11-032, and allows for further unexpected discoveries.

\par The remainder of the paper is organized into 3 sections. In section 2 we analyze the new Chandra observation and search for evidence of either a dual AGN (via a spatial analysis) and/or a binary AGN (via a spectral analysis); in section 3 we incorporate archival multi-wavelength coverage into our analysis to investigate sources resolved in the Chandra dataset; and in section 4 we summarize our findings. Throughout the paper we assume a $\Lambda$CDM universe, where $H_{0}=69.6$ km s$^{-1}$, $\Omega_{M}=0.286$, and $\Omega_{\Lambda}=0.714$.

\begin{figure}
    \centering
    \includegraphics[width=\linewidth]{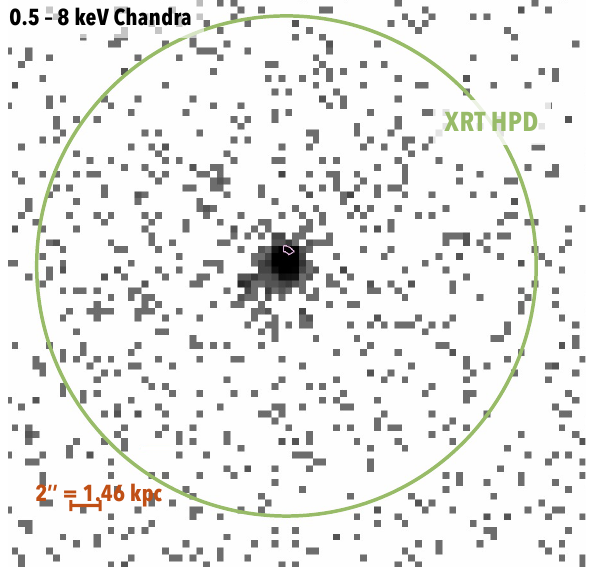}
    \caption{The 0.5$-$8 keV Chandra observation of MCG+11-11-032. The X-ray image has been binned to Chandra's native pixel resolution. In green, we show the Swift XRT half-power diameter at 1.5 keV (18\arcsec). By-eye, the observation shows extended X-ray emission associated with a secondary point source on scales $>2$\arcsec\. The extension sits at a separation from the central AGN that is larger than the known artifact in Chandra's PSF (which sits within the central arcsecond, shown in pink). North is up and East is to the left, and a 2\arcsec\ bar (1.46 kpc at $z=0.0362$) is shown for scale. }
            \label{fig:Xrayobs}
\end{figure}

\section{X-ray Analysis and Results} \label{sec:method}
MCG+11-11-032 was targeted by Chandra in a Cycle 23 proposal (PI: Civano, ID: 23700236). The AGN was observed for 35 ks on 03-07-2023 (observation ID: 25367). The source was observed on-axis and placed on the back-illuminated S3 chip of the ACIS detector. The exposure time was set to achieve at least 1300 (3300) counts in the 0.5$-$8 keV band in the faint (bright) state with the goal of constraining both of the putative Fe lines at $>3\sigma$. To avoid pile-up, which could compromise the goal of separating two nuclei if the source is in the bright state, MCG+11-11-032 was observed in a sub-array mode (standard 1/8 to reduce the CCD frame time to 0.4 sec). 

We follow the standard data reduction for point source observations, using Chandra Interactive Analysis of Observations software ({\tt CIAO}) v4.12 \citep{Fruscione2006}. The first step is to correct for astrometry, cross-matching the Chandra-detected point-like sources with the SDSS Data Release 18 (SDSS DR18) catalog. The Chandra sources used for cross-matching are detected by running {\tt wavdetect} on the reprocessed level-2 event file. We require the observation to have a minimum of 3 matches with the SDSS DR18, and each matched pair to be less than 2\arcsec\ from one another.  Because the observation for MCG+11-11-032 was taken in a sub-array, we find that there are not enough X-ray point sources to match with the SDSS DR18 catalog.  However, we note that all Chandra observations undergo aspect corrections via the AGASC catalog\footnote{https://cxc.harvard.edu/cgi-gen/cda/agasc/agascInterface.pl}, with current radial astrometric errors for on-axis sources of $\approx$ 0.79\arcsec\footnote{https://cxc.harvard.edu/cal/ASPECT/celmon/}. The lack of additional astrometric corrections have no effect on our X-ray data analysis: if a dual AGN, the locations of each putative AGN will be measured relative to the primary X-ray point source; if a binary AGN, relative astrometric shifts will not affect our X-ray spectral analysis. Lastly, we find the background flaring contribution to be negligible, with no time interval containing a background rate 3$\sigma$ above the mean level.  

To quantitatively determine the presence of one or multiple point sources, we analyze the Chandra data with  \BAYMAX (see, e.g., \citealt{Foord2019, Foord2020, Sandoval2024}). \BAYMAX (i) takes calibrated Chandra events and compares them to the expected distribution of counts for single versus dual point source models; (ii) calculates a Bayes factor to determine which model is preferred ($\logBF$); (iii) calculates likely values for angular separation ($r$) and count ratio ($f$, ratio between the secondary and primary source); (iv) probabilistically assigns sources counts to each model component (iv) and fits spectra to each component, allowing for estimates of the flux. In addition to each point source component, all models includes a background component, which is energy independent and assumed to be uniformly distributed. For bright sources, source counts and fluxes obtained by \BAYMAX match the expected output using standard {\tt CIAO} tools \citep{Foord2019, Foord2020}. However, for low-count and/or closely separated components, the statistical spectral analysis carried out by \BAYMAX allows for a more robust measure of X-ray properties. 

For each source, the nested sampling technique that calculates the Bayes factors returns a statistical error bar. In past analyses, we have found that the statistical error bars are consistent with the 1$\sigma$ spread in the $\logBF$ values when running \BAYMAX 100 times on a given observation \citep{Foord2019, Foord2020, Foord2021}. In the following section, we quote a 3$\sigma$ error on the $\logBF$ value using the statistical errors returned by \BAYMAX. For datasets in a similar region of parameter space as our Chandra observation (on-axis observations with over $700$ counts between 0.5$-$8 keV), we have previously quantified a ``strong'' Bayes factor by running \BAYMAX on simulations of single AGN and calculating the false positive rate below various thresholds; all simulations were found to have $\log{BF}<0.54$ in favor of the dual point source, and thus we use this threshold to define a strong $\log{BF}$ value in favor of the dual point source model \cite{Foord2019} With over 700 counts, we can expect to be sensitive to detecting dual AGN down to angular separations of 0.5\arcsec\ at count-ratio values $f\ge0.1$. For detailed information on the statistics behind the code, we refer the reader to \cite{Foord2019}. 

\subsection{Dual AGN Hypothesis: Spatial Analysis of Chandra Dataset}
By-eye, the Chandra observation shows extended X-ray emission near the location of MCG+11-11-032, separated on scales $>$2\arcsec. In Figure~\ref{fig:Xrayobs}, we show the 0.5$-$8 keV Chandra observation. This separation is larger than the known placement of the artifact in Chandra's point-spread function, and has a different position angle with respect to the feature's expected location (see Fig. 1). We analyze the 0.5$-$8 keV counts with \BAYMAX, using a $20\arcsec \times 20\arcsec$ sky region centered on the nominal X-ray coordinates of the AGN (corresponding to a $13~\text{kpc}\times13~\text{kpc}$ box at $z=0.0362$). Similar to previous analyses using {\tt BAYMAX} \citep{Foord2020, Foord2021, Sandoval2024}, the single point source model's prior distributions include the source's sky coordinate, $\mu$, and the logarithm of the background fraction, $\log f_{\text{bkg}}$. The dual point source model's prior distributions include the sky coordinates for each source, $\mu_1$ and $\mu_2$, the logarithm of the background fraction, $\log f_{\text{bkg}}$, and the logarithm of the count ratio, $\log f$. For both models, $f_{\text{bkg}}$ is defined as the ratio of the number of counts associated with the background versus the number of counts associated with all point sources. For the dual point source model, $f$ is defined as the ratio of the number of counts between the secondary and primary X-ray point source. For all priors, we use the standard ranges and distributions, as most recently detailed in \cite{Sandoval2024}. All errors evaluated in the following sub-section are done at the 95\% confidence level. 

\begin{figure*}
    \centering
    \includegraphics[width=0.9\linewidth]{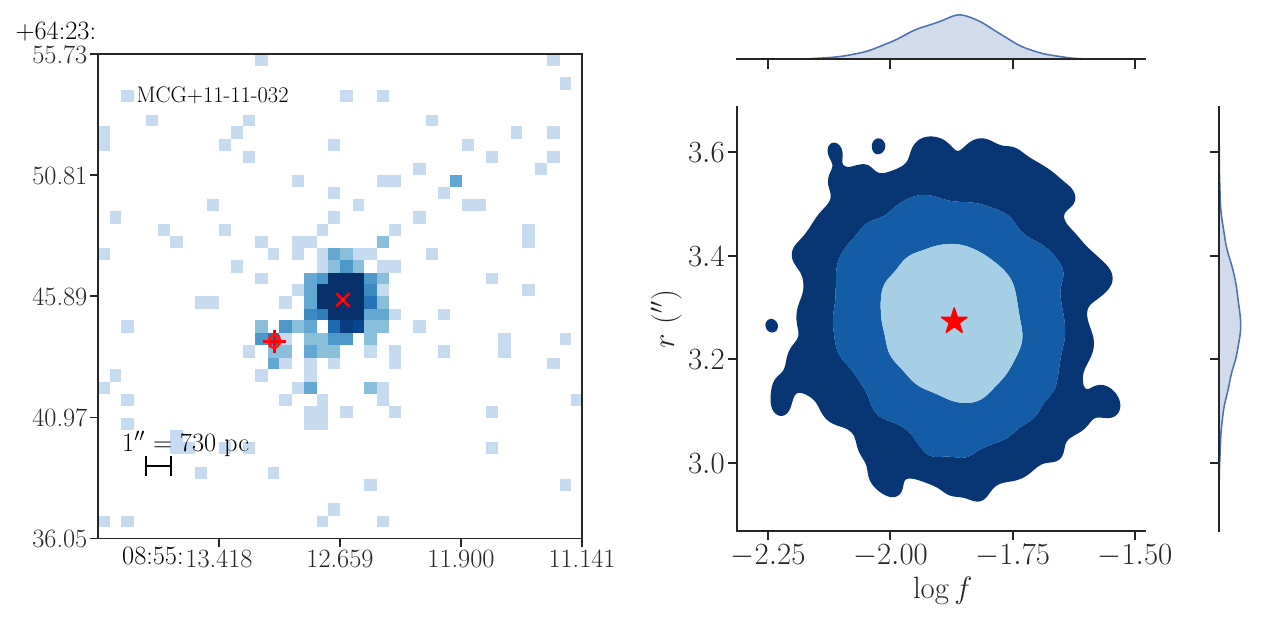}
    \caption{Results from \BAYMAX showing the best-fit locations for the primary and secondary X-ray point sources over the $0.5$--$8$ keV dataset (\emph{left}) and the joint posterior distributions for $r$ and $\log{f}$ (\emph{right}). In the left panel, we plot the 95\% confidence intervals in red lines for the best-fit location, which are smaller than the symbol for the primary point source. The X-ray image has been binned to Chandra's native pixel resolution; North is up and East is to the left. In the right panel, the 68\%, 95\%, and 99.7\% confidence intervals for $r$ and $\log{f}$ are shown from light to dark blue contours. We denote the location of the median of the posterior distributions with a red star.}
    \label{fig:BAYMAXresults}
\end{figure*}

Analysis of the dataset with \BAYMAX yields a Bayes Factor $\logBF = 37.87 \pm 1.95$ in favor of the dual point source model. The position of the secondary X-ray point source coincides with the position of the extended emission, and the best-fit separation and count ratio returned by \BAYMAX are $r=3.27\arcsec_{-0.27}^{+0.32}$ (approximately 2.4 kpc at $z=0.0362$) and $\log{f}=-1.86_{-0.28}^{+0.28}$. The calculated position angle between the primary AGN (hereafter MCG+11-11-032 X1, for simplicity) and the secondary X-ray point source (hereafter MCG+11-11-032 X2, for simplicity) is $\text{PA}_{\text{X}}=149\pm{4}^{\circ}$. This position angle is approximately perpendicular to the position angle of the galaxy's plane and the value where the two resolved narrow emission line peaks were previously determined to be maximally separated (PA$_{\text{opt}}$=46$^{\circ}$; \citealt{Comerford2012}) Given that the best-fit coordinates of the two X-ray sources correspond to a separation over 4 times as large, and a position angle that is perpendicular to the values reported in \cite{Comerford2012}, it is likely that this source is not the origin for the observed double-peaked [\ion{O}{3}] emission lines. In Figure~\ref{fig:BAYMAXresults} we show the results from \BAYMAX, including the best-fit locations of the each X-ray point source on the Chandra image, and the joint posterior distributions for $r$ and $\log{f}$.

\begin{table*}
\caption{X-ray Spectral Properties of MCG+11-11-032}
\label{tab:spectralmodels}
\setlength\tabcolsep{4.5pt}
\begin{tabular*}{\textwidth
}{cccccccc}
	\hline 
	\hline 
	\multicolumn{8}{c}{X-ray Spectral Model Comparison: MCG+11-11-032 X1} \\
	\hline \\ [-1.7ex]
	\multicolumn{1}{c}{Model} &
	\multicolumn{1}{c}{$\Gamma$} &
	\multicolumn{1}{c}{$N_{H}$ (10$^{22}$ cm$^{-2}$)} &
	\multicolumn{1}{c}{$E$ (keV)} &
	\multicolumn{1}{c}{$EW$ (eV)} &
 	\multicolumn{1}{c}{$F_{2-10~\mathrm{keV}}$ ($10^{-12}$ erg s$^{-1}$ cm$^{-2}$)} &
	\multicolumn{1}{c}{$L_{2-10~\mathrm{keV}}$ ($10^{42}$ erg s$^{-1}$)}  & \multicolumn{1}{c}{$C_{\mathrm{stat}}$/\textit{dof}/$n_{\text{fp}}$} \\
	\multicolumn{1}{c}{(1)} & \multicolumn{1}{c}{(2)} & \multicolumn{1}{c}{(3)} & \multicolumn{1}{c}{(4)}  & \multicolumn{1}{c}{(5)}  & \multicolumn{1}{c}{(6)} & \multicolumn{1}{c}{(7)} & \multicolumn{1}{c}{(8)} \\ [1.ex]
	\hline \\ [-2.5ex]
	M1 & $1.02_{-0.02}^{+0.68}$ & $11.44_{-0.15}^{+4.15}$ & $6.33_{-0.17}^{+0.26}$ & $66_{-58}^{+60}$ & $3.38^{+0.09}_{-0.41}$ & $9.68_{-1.04}^{+0.25}$ & 513/505/7  \\ [0.4ex]
    \hline \\ [-2.7ex]
    M2 & $1.03_{-0.3}^{+0.66}$ & $11.49_{-0.13}^{+3.89}$ & $6.33_{-0.22}^{+0.53}$ &  $23_{-20}^{+83}$ & $3.38_{-0.41}^{+0.08}$ & $9.68_{-1.03}^{+0.23}$ & 513/503/9  \\ [0.4ex]
    & & & $6.34_{-0.25}^{+0.42}$ & $40_{-36}^{+72}$ & & \\
    \hline \\ [-2.7ex]
    M3 & $1.02_{-0.02}^{+0.62}$ & $11.67_{-0.45}^{+3.51}$ & 6.16 (fixed) & $2_{-2}^{+5}$ & $3.40_{-0.41}^{+0.07}$ & $9.73_{-1.02}^{+0.27}$ & 521/506/7  \\ [0.4ex]
    & & & 6.56 (fixed) & $5_{-3}^{+3}$ & & \\
    \hline \\ [-2.7ex]
    M4 & 1.23$_{-0.07}^{+0.72}$ & 12.44$_{-0.75}^{+4.87}$ & 6.33$_{-0.12}^{+0.21}$ & $75_{-53}^{+56}$  & $3.27_{-0.39}^{+0.16}$ & $9.43_{-1.10}^{+0.43}$ & 514/504/9 \\ [0.4ex]
     & & & $7.56_{-0.31}^{+0.18}$ & $114_{-95}^{+161}$ &  &  \\ [0.4ex]
	\hline 
	\multicolumn{8}{c}{\BAYMAX Best-fit Spectral Parameters: MCG+11-11-032 X1 and X2} \\
	\hline \\ [-1.7ex]
	\multicolumn{1}{c}{Source} &
	\multicolumn{1}{c}{$\Gamma$} &
	\multicolumn{1}{c}{$N_{H}$ (10$^{22}$ cm$^{-2}$)} &
	\multicolumn{1}{c}{$E$ (keV)} &
	\multicolumn{1}{c}{$EW$ (eV)} &
 	\multicolumn{1}{c}{$F_{2-10~\mathrm{keV}}$ ($10^{-12}$ erg s$^{-1}$ cm$^{-2}$)} &
	\multicolumn{1}{c}{$L_{2-10~\mathrm{keV}}$ ($10^{42}$ erg s$^{-1}$)} &
	\multicolumn{1}{c}{$HR$}  \\
	\multicolumn{1}{c}{(1)} & \multicolumn{1}{c}{(2)} & \multicolumn{1}{c}{(3)} & \multicolumn{1}{c}{(4)}  & \multicolumn{1}{c}{(5)}  & \multicolumn{1}{c}{(6)} & \multicolumn{1}{c}{(7)} & \multicolumn{1}{c}{(8)} \\ [1.ex]
	\hline \\ [-2.5ex]
	X1 & $1.38_{-0.05}^{+0.06}$ & $13.93_{-0.35}^{+0.43}$ & $6.37_{-0.02}^{+0.01}$ & $55_{-5}^{+6}$ & $2.87_{-0.03}^{+0.02}$ & $8.30_{-0.07}^{+0.06}$ & $0.89_{-0.03}^{+0.02}$ \\ [0.4ex]
 	 & & & $7.55_{-0.02}^{+0.01}$ & $162_{-21}^{+17}$ & & & \\ [0.4ex]
    \hline \\ [-2.7ex]
    X2 & 1.8 (fixed) &  $9.38_{-2.94}^{+10.30}$ & & & $3.03_{-0.86}^{+0.69} \times 10^{-2}$ & $1.52_{-0.48}^{+0.96}\times10^{-1}$ & $0.88_{-0.09}^{+0.11}$ \\ [0.4ex]
 
    \hline
    \hline
\end{tabular*}
\textit{Top:} Columns: (1) Spectral model used; (2)  the best-fit spectral index; (3) the best-fit extragalactic column density; (4) the best-fit central line energy for {\tt zgaus} component; (5) the equivalent width measured for the {\tt zgaus} component; (6) the measured $0.5$--$8$ keV flux, in units of 10$^{-12}$ erg s$^{-1}$ cm$^{-2}$; (7) the rest-frame $2$--$10$ keV luminosity in units of 10$^{42}$ erg s$^{-1}$; (8) the $C_{\mathrm{stat}}$, degrees of freedom (\textit{dof}), and $n_{\text{fp}}$ values associated with each model and used to quantify the best-fit (see text for more details). \textit{Bottom:} Columns: (1) Source name; (2)  the assumed or best-fit spectral index; (3) the best-fit extragalactic column density; (4) the best-fit central line energy for {\tt zgaus} component; (5) the best-fit equivalent width measured for the {\tt zgaus} component; (6) the measured $0.5$--$8$ keV flux, in units of 10$^{-12}$ erg s$^{-1}$ cm$^{-2}$; (7) the rest-frame $2$--$10$ keV luminosity in units of 10$^{42}$ erg s$^{-1}$; (8) the hardness ratio (see text for definition). Each best-fit value is defined as the median of the full distribution, see text for more details. For MCG+11-11-032 X2, Column 7 represents the unabsorbed luminosity, calculated using the {\tt XSPEC} model component {\tt cflux}.
\end{table*}

\subsection{Binary AGN Hypothesis: Spectral analysis of Chandra Dataset}
The spectral analysis is performed by using the XSPEC 12.13.1 package \citep{Arnaud1996}. We use Poisson likelihood statistics ({\tt cstat}) to compare and determine the best-fit model \citep{Cash1979}.  Specifically, when comparing two models, a statistically significant improvement in the fit is defined when $\Delta C_{\mathrm{stat}}>$ $\Delta n_\text{fp} \times$2.71 (where $\Delta n_\text{fp}$ represents the difference in number of free parameters between the models; \citealt{Tozzi2006, Brightman2012}), corresponding to a fit improvement with 90\% confidence (\citealt{Brightman2014}).
Error bars quoted in the following section are calculated with Monte Carlo Markov Chains via the XSPEC tool {\tt chain} at the 90\% confidence level for each parameter of interest. More details about the best-fit X-ray spectral parameters for each model described below can be found in Table~\ref{tab:spectralmodels}.

We analyze the photons extracted within a 2\arcsec\ radius circular region centered on the X-ray coordinates of MCG+11-11-032 and corresponding to 95\% of the encircled energy radius at 1.5 keV for ACIS-S.  For our background extraction, we use a source-free annulus with an inner radius of 10\arcsec\ and outer radius of 20\arcsec. We calculate a count rate of approximately $6.31 \times 10^{-2}$ cts/s, representing a state in-between the low- and high-states observed with XRT. For the most precise comparison to the results presented in \cite{Severgnini2018}, we adapt their best-model: {\tt tbabs*(ztbabs*zpowerlw+pexrav+zgaus)}, hereafter Model 1. This model represents an intrinsically absorbed powerlaw ({\tt ztbabs*zpowerlw}) plus a continuum reﬂection component ({\tt pexrav}, \citealt{Magdziarz1995}), and a narrow emission-line component ({\tt zgaus}) that accounts for possible Fe K$\alpha$ emission.  The {\tt pexrav} component was originally added due to both soft- ($<2$ keV) and the hard-energy ($>6$ keV) residuals when modeling the spectrum with an absorbed powerlaw (suggesting the presence of a reflection component that may be due to, e.g.,  circumnuclear material).

We fix the photon index of the powerlaw, $\Gamma$, to the photon index of the reflection component. We set the initial value of $\Gamma$ to 1.8, and allow the value to vary between 1 and 3 \citep{Corral2011,Yan2015}. We set the {\tt pexrav} reflection scaling factor ({\tt rel}) to a value of -1, such that it represents the reflection component only. We set the initial line energy value for the {\tt zgaus} component to the best-fit value found in \cite{Severgnini2018} ($E = 6.18$ keV), allowing the value to vary in the 6$-$7 keV energy range. Interestingly, \cite{Severgnini2018} find the emission line component is signiﬁcantly lower (at 97\% conﬁdence level) than the nominal rest-frame value for Fe K$\alpha$ near $\sim$6.4 keV. This energy value has no clear association with well-known and expected transitions in the X-ray band. Following \cite{Severgnini2018}, we also freeze the sigma value to 50 eV. We find best-fit values for the Gaussian emission component of $E=6.33_{-0.17}^{+0.26}$ keV with equivalent width $EW = 66_{-58}^{+60}$ eV. The best-fit properties of the emission line include a higher-energy (that is consistent with the nominal rest-frame value for Fe K$\alpha$ near $\sim$6.4 keV), and lower equivalent width than presented in \cite{Severgnini2018}. However, we find a power-law slope $\Gamma$ that is pegged at, and consistent with, the lower limit of $1$.

We test for the presence of two emission lines by adding a second ({\tt zgaus}) component to our model. We set the initial line energy values for both components to the best-fit values found in \cite{Severgnini2018} when fitting the XRT data with two Gaussian components (their model 4; $E_{1} = 6.16$ keV and $E_{2} = 6.56$ keV). We allow the values to vary in the 6$-$7 keV energy range and freeze the sigma values to 50 eV (hereafter Model 2). We find that this fit does not result in a statistically significant improvement. Futhermore, the two emission lines are shifted from their initial values and centered at nearly the same value for the single Gaussian component in Model 1 ($E_{1} = 6.33_{-0.22}^{+0.53}$ keV and $E_{2} = 6.34_{-0.25}^{+0.42}$ keV), but with approximately half the normalization values. Similar to Model 1, the best-fit value for $\Gamma$ is pegged at, and consistent with, our lower-boundary of 1. The degeneracy between the results of Model 1 and Model 2 is clearly seen when comparing the flux and luminosity values, which are nearly identical. 

In an attempt to quantify the statistical significance of a model with two distinct emission lines, we fit the data using the same components as in Model 2, but freeze the line energy values of both {\tt zgaus} components to $E_{1} = 6.16$ keV and $E_{2} = 6.56$ keV (hereafter Model 3). This fit does not result in a statistically significant improvement. In particular, the normalizations of both emission lines are shifted to values a factor of 10 smaller than Model 2, such that the equivalent widths of each line are consistent with, or close to, 0. Similar to both Model 1 and Model 2, the best-fit value for $\Gamma$ is pegged at, and consistent with, a value of 1. 

\begin{figure}
    \centering
    \includegraphics[width=\linewidth]{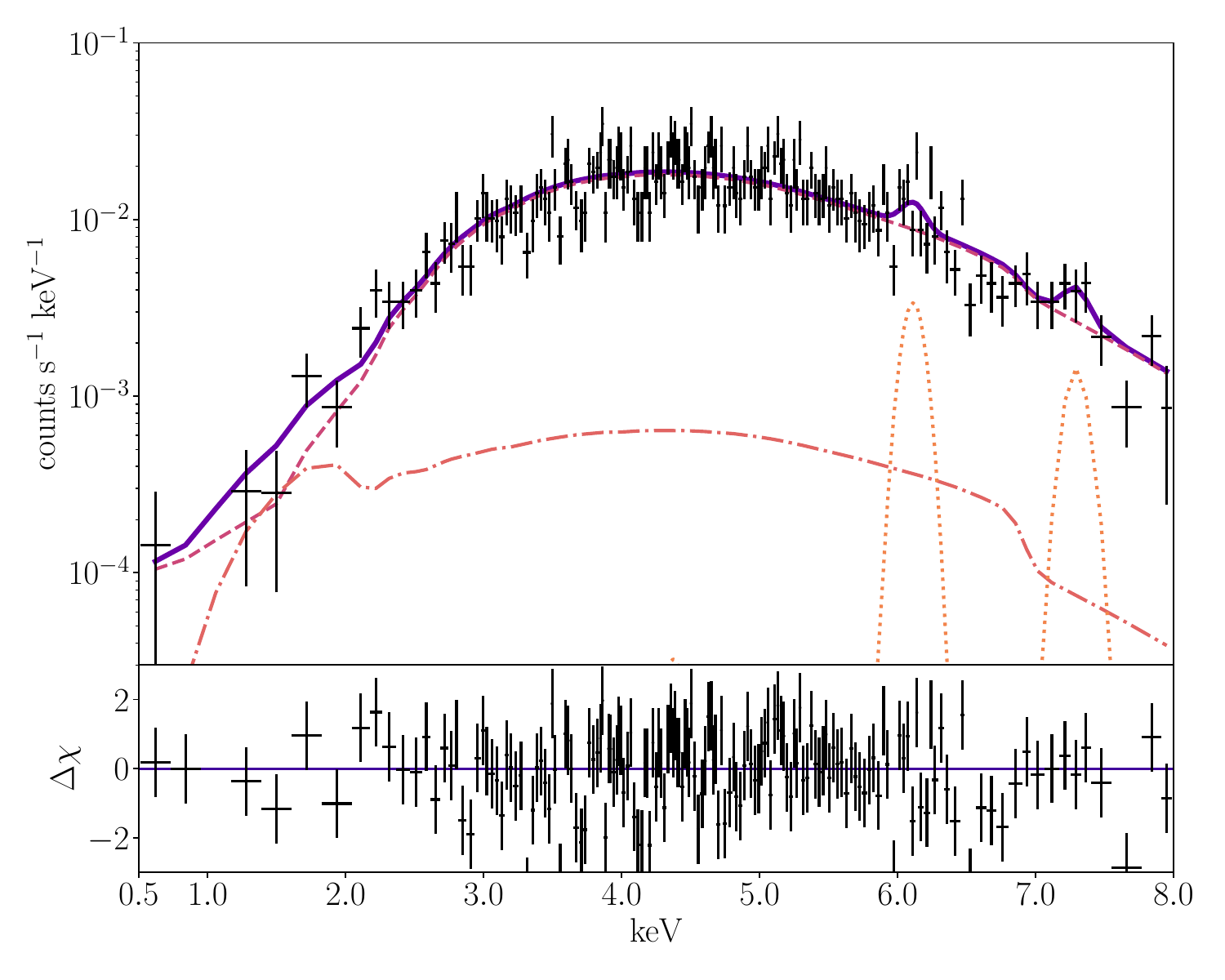}
    \caption{\emph{Top:} Spectrum of the central AGN shown in black, with the best-fit model overplotted in dark purple. We find that Model 4 results in the best-fit, defined as: {\tt tbabs*(ztbabs*zpowerlw+pexrav+zgaus+zgaus)}. We plot each component individually, with {\tt zpowerlw} in pink dashed line; {\tt pexrav} in orange dot-dashed line; and both {\tt zgaus} components in gold dotted line. We allow each {\tt zgaus} component to vary between 6$-$7 keV and 7$-$8 keV, and find best-fit values of: $E_{1}=6.33_{-0.12}^{+0.21}$ keV with $EW = 75_{-53}^{+56}$ eV, and $E_{2}=7.56_{-0.31}^{+0.18}$ keV with $EW = 114_{-95}^{+161}$ eV. We find no evidence of double-peaked narrow Fe K$\alpha$ lines with an orbital velocity offset of $\Delta v \approx 0.06$ in the X-ray spectrum of MCG+11-11-032. The emission line at $\approx 7.5$ keV can be possibly associated with either Fe K$\beta$, Ni K$\alpha$, or a combination of the two. \emph{Bottom:} Residuals between data and model, defined as (data-model)/error where error is calculated as the square root of the model predicted number of counts.} 
    \label{fig:bestfitspec}
\end{figure}

Residuals between the data and the best-fit spectrum for all of our models show an excess of emission between $7-8$ keV. We fit the data using the same components in Model 2,  but with initial line energy values for each {\tt zgaus} component to $E_{1} = 6.33$ keV (allowed to vary between 6$-$7 keV) and $E_{2} = 7.5$ keV (allowed to vary between 7$-$8 keV). We freeze the sigma values to 50 eV (hereafter Model 4). We find best-fit values for the Gaussian emission components of $E_{1}=6.33_{-0.12}^{+0.21}$ keV with an equivalent width $EW = 75_{-53}^{+56}$ eV, and $E_{2}=7.56_{-0.31}^{+0.18}$ keV with $EW = 114_{-95}^{+161}$ eV. With respect to Model 1, this fit does not result in a statistically significant improvement given the additional number of free parameters; however, Model 4 is the only model that results in a $\Gamma$ value that is not pegged at the boundary value of 1. Thus, we conclude that Model 4 best describes the X-ray emission associated with MCG+11-11-032. We find no evidence of double-peaked narrow Fe K$\alpha$ lines with an orbital velocity offset of $\Delta v \approx 0.06$ in the X-ray spectrum of MCG+11-11-032. The emission line at $\approx 7.5$ keV can be possibly associated with either Fe K$\beta$, Ni K$\alpha$, or a combination of the two. In Figure~\ref{fig:bestfitspec} we show our best-fit model (Model 4) overplotted on the data. 

\section{Discussion} \label{sec:results}
Regarding the dual AGN hypothesis, we find that the Chandra observation has spatially extended X-ray emission (MCG+11-11-032 X2) approximately 3\arcsec\ from the central AGN (MCG+11-11-032 X1). However, we find that the position angle between MCG+11-11-032 X1 and X2 is perpendicular to the previously calculated position angle between resolved double-peaked narrow [\ion{O}{3}] emission lines in the galaxy \citep{Comerford2012}. Regarding the binary AGN hypothesis, we find no strong evidence of double-peaked Fe K$\alpha$ emission in the Chandra spectrum. In the following section, we further discuss possible origins of MCG+11-11-032 X2 and how it may have contaminated previous analyses searching for evidence of a binary AGN.

\subsection{Possibility of Outflows in MCG+11-11-032}
Importantly, our initial analysis with \BAYMAX models the X-ray emission for any detected component as a point source, as expected for AGN. Some of the emission asociated with MCG+11-11-032 X2 may be due to the presence of an outflow. AGN outflows can interact with the host galaxy and surrounding medium, resulting in large-scale ionization detectable in both X-ray and optical wavelengths. Specifically, Seyfert 2 galaxies have shown soft ($<3$ keV) extended X-ray emission that is spatially correlated with \ion{O}{3} emission, indicating that photoionization might be the common origin for these phenomena (\citealt{Levenson2006, Bianchi2006, Bianchi2010, Travascio2021}).

The observations of MCG+11-11-032 from the Blue Spectrogram on the MMT 6.5 m telescope were re-analyzed in \cite{Nevin2016}, as part of a study on 71 double-peaked narrow emission line sources. A kinematic classification scheme was used to classify each spectra in order to determine the origin of the double-peaked [\ion{O}{3}] emission. The majority (86\%) of the sample was found to have [\ion{O}{3}] profiles well-described by moderate-luminosity AGN outflows. However, MCG+11-11-032 was classified as ``Ambiguous'' (the smallest classification group, only 8\% of the sample), given that the line-of-sight velocity and the dispersion of [\ion{O}{3}] were not unambiguously associated with outflows ($<400$ km s$^{-1}$ and $<500$ km s$^{-1}$, respectively), and that the [\ion{O}{3}] emission was found to be misaligned with the galaxy plane. The extension of the [\ion{O}{3}] emission was found to have a position angle of PA$_{\text{opt}}= 137^{\circ}$, similar to the position angle determined from the X-ray analysis (PA$_{X}=149\pm4^{\circ}$).

\begin{figure*}
        \centering
        \includegraphics[width=\linewidth]{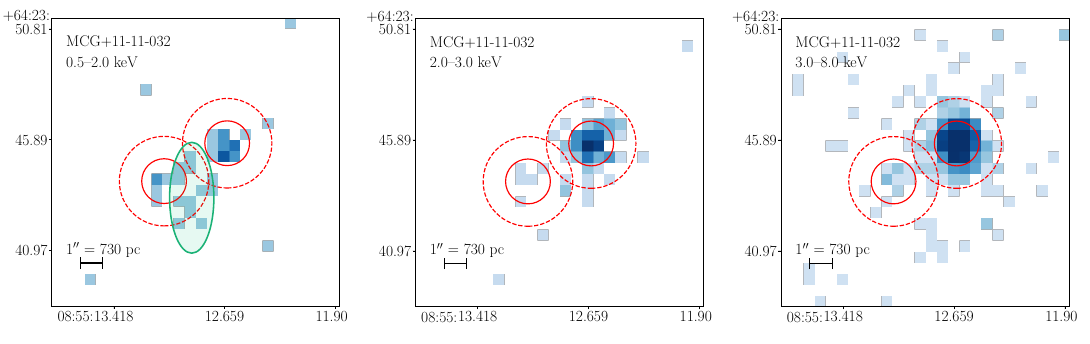}
        \caption{X-ray data of MCG+11-11-032 in different energy bins: 0.5$-$2 keV (left), 2$-$3 keV (center), and 3$-$8 keV (right). The X-ray images are binned to Chandra's native pixel resolution. In solid (dashed) red lines we plot $1\arcsec$ ($2\arcsec$) radius circles centered on the best-fit location for the primary and secondary as determined by \BAYMAX when analyzing the full 0.5$-$8 keV dataset (see Section 2.1). In the lowest-energy bin we find evidence for soft extended emission on $\approx4\arcsec$~scales, shown with a green ellipse. The source is dominated by photons with energies $<2$~keV, as we find no evidence for extended emission in the 2$-$3 keV or 3$-$8 keV bands. By-eye, a compact source is observed at all energy bins, consistent with the location of the secondary point source as determined by \BAYMAX. The majority of counts are contained within 1$\arcsec$, as expected for a point source. Furthermore, the ratio of counts within a 1$\arcsec$ radius around the coordinates of the secondary and primary is consistent across all energy bins ($\approx 0.01$) and with our previous results obtained using \BAYMAX ($\log{f} = -1.86$). The exception is in the 0.5$-$2 keV range (0.47), where contamination from the extended emission is expected.}
        \label{fig:energybreakup}
\end{figure*}

We evaluate the Chandra observation for the presence of outflows by analyzing 3 energy bins: 0.5$-$2  keV (where emission associated with outflows may dominate); 2$-$3 keV; and 3$-$8 keV. In Fig.~\ref{fig:energybreakup} we show the filtered Chandra data. In the lowest-energy bin, we find evidence for extended soft emission (on $\sim$4\arcsec\ scales, or 2.9 kpc at $z=0.0362$). The source appears to be dominated by $<2$ keV photons, as there is no evidence for extended emission in the 2$-$3 keV or 3$-$8 keV bands. In all energy bands there exists a compact source at the coordinates of MCG+11-11-032 X2, as previously returned by \BAYMAX. Within a 1\arcsec\ extraction region centered on these coordinates we calculate a hardness ratio ($HR$), defined as $(H - S)/(H + S)$ where $H$ and $S$ are the number of hard (2$-$8 keV) and soft (0.5$-$2 keV) X-ray counts. The $\emph{HR}$ value for a given source is independent of the model used to describe the source, and can be used to identify characteristics of a given emitter in the low-count regime. We find HR=$0.48$, consistent with the values observed for type-2 AGN across a range of expected extragalactic column densities (\citealt{Hasinger2008}). Due to contamination from the soft extended region, the true value is likely higher.

We calculate the ratio of counts within a 1\arcsec\ radius circle centered on the coordinates of MCG+11-11-032 X2 and MCG+11-11-032 X1, as a function of energy. We find the ratios consistent across all energy bins ($\approx 0.1$), and with our previous results obtained using \BAYMAX ($\log{f} = -1.86$). The exception is in the 0.5$-$2 keV range ($\approx 0.47$), where contamination from the extended component is expected. Given the low number of counts associated with the soft extended component ($\approx$ 12), it is challenging to constrain other properties, such as the position angle and spectrum. However, the soft extended X-ray component and the double-peaked narrow [\ion{O}{3}] detected in MCG+11-11-032 reflect that large-scale emission around the nucleus of galaxy MCG+11-11-032 is multi-faceted and may be partially composed of low-level outflows (although other X-ray contaminants that can contribute soft extended emission exist, including, e.g., star formation). We note that the kinematics powering the double-peaked [\ion{O}{3}] emission in ``ambiguous'' sources is unclear: their spectra can be explained by many scenarios including a counter-rotating disk caused by a merger, an outflow, an inflow, or some combination \citep{Nevin2016}.

To test the legitimacy of the assumption that MCG+11-11-032 X2 is a hard compact point source, separate from the soft extended X-ray emission observed near the nucleus, we carry out two analyses. The first is re-running our analysis with \BAYMAX as presented in Section 2.1, but on only the hardest photons in the observation (3$-$8 keV) to evaluate if the dual point source model remains favored over the single point source model; although we decrease the number of counts in the analysis by approximately 10\% (and thus lose statistical power), this energy range eliminates the majority of contamination associated with diffuse emission (see Fig.~\ref{fig:energybreakup}). Our first analysis yields $\log{\BF} = 9.38 \pm 1.95$ in favor of the dual point source model. The position of the secondary X-ray point source is consistent with results presented in Section 2.1, while the count ratio is marginally lower ($\log{f} \approx$-2.1; as expected due to less contamination from the extended soft region). In Fig.~\ref{fig:appendix} we show the results from the 3$-$8 keV analysis with \BAYMAX, where each count has been probabilistically assigned to a model component.

The second is re-running our analysis on the 0.5$-$8 keV photons with \BAYMAX using updated models to quantify whether the dual point source model remains favored when including a diffuse emission region near the location of MCG+11-11-032 X2 (modeled as a component with a spatially uniform distribution of photons). In particular, it is possible that a spatially uniform component with a high count-rate sitting amongst a spatially uniform background with a lower count-rate can be mistaken for a secondary point source. We update our single and dual models to include an additional high-count and spatially uniform component. This component is defined within an elliptical region that covers the majority of the extended emission (as determined in the 0.5$-$2 keV range), with a semi-major and semi-minor axis of 5\arcsec\ and 2\arcsec\, respectively (see Fig.~\ref{fig:energybreakup}). The region overlaps within 1\arcsec\ of the best-fit location for MCG+11-11-032 X2, as determined in Section 2.1. Within this region, the background component is replaced with a spatially uniform emission component that has a different count-rate than the background. This model component is parameterized by $f_{\text{diff}}$, which represents the fraction of counts in the region that are associated with the diffuse high-count component. We include this parameter in both the single and double point source models, with the goal of quantifying if a single point source plus a region of high count-rate diffuse emission can statistically describe the data better than a model that includes a second point source. We emphasize that the because the majority of the extended X-ray emission is observed below 2 keV (see Fig.~\ref{fig:energybreakup}), this analysis likely overestimates the contribution of the photons associated with the extended component. We find a Bayes factor value of $\log{\BF} = 28.6 \pm 2.2$ in favor of the dual point source model. In Fig.~\ref{fig:appendix} we show the results from this analysis, where each count has been probabilistically assigned to each model component. Even with an additional high-count component, we find strong evidence for the presence of a point source.

Thus, although both the X-ray and optical wavebands \citep{Nevin2016} suggest that MCG+11-11-032 has low-level outflows, we find evidence of an additional compact and hard point source that sits approximately 3.27\arcsec\ from the nucleus in MCG+11-11-032. Ground-based integral field unit spectroscopy (IFU) observations will allow for a much more detailed mapping of the optical ionization around the nucleus of MCG+11-11-032, including extent, geometry, and spectrum.

\subsection{X-ray Spectral Analysis of MCG+11-11-032 X2}
Assuming that MCG+11-11-032 X2 can be well-represented as a point source, we investigate the origin of the emission by analyzing its X-ray spectrum. As outlined in \cite{Foord2020}, \BAYMAX{} can be used to carry out a spectral analysis of each detected point source component. We use our best-fit results from the full energy band (0.5$-$8 kev) as presented in Section 2.1, with the caveat that we expect low-level contamination (fewer than 2 counts) in both the spectra of MCG+11-11-032 X1 and MCG+11-11-032 X2 at photon energies $<2$ keV from the extended soft source.

We create 100 spectral realizations of each component by probabilistically sampling from the full distribution of counts. Each spectral realization is associated with [$\mu_{1}$, $\mu_{2}$, $f_{\text{bkg}}$, $f$] values that are drawn from the posterior distributions of the dual point source model. Each count is then assigned to a specific model component (i.e, the primary, secondary, or background), based on the respective probabilities of being associated with each component. On average, the secondary point source's spectrum only has 27 counts (as compared to the primary's average of 1965 counts), which limits the complexity of our spectral model. We fit each spectral realization of both point source components with XSPEC; all of the spectral realizations of the primary point source (MCG+11-11-032 X1) are fit with the same constraints used in Model 4, while the secondary point source's (MCG+11-11-032 X2) spectral realizations are fit with an absorbed powerlaw ({\tt tbabs*(ztbabs*zpowerlw)}), with a fixed photon index $\Gamma=1.8$. 

\begin{figure*}
        \centering
        \includegraphics[width=0.9\linewidth]{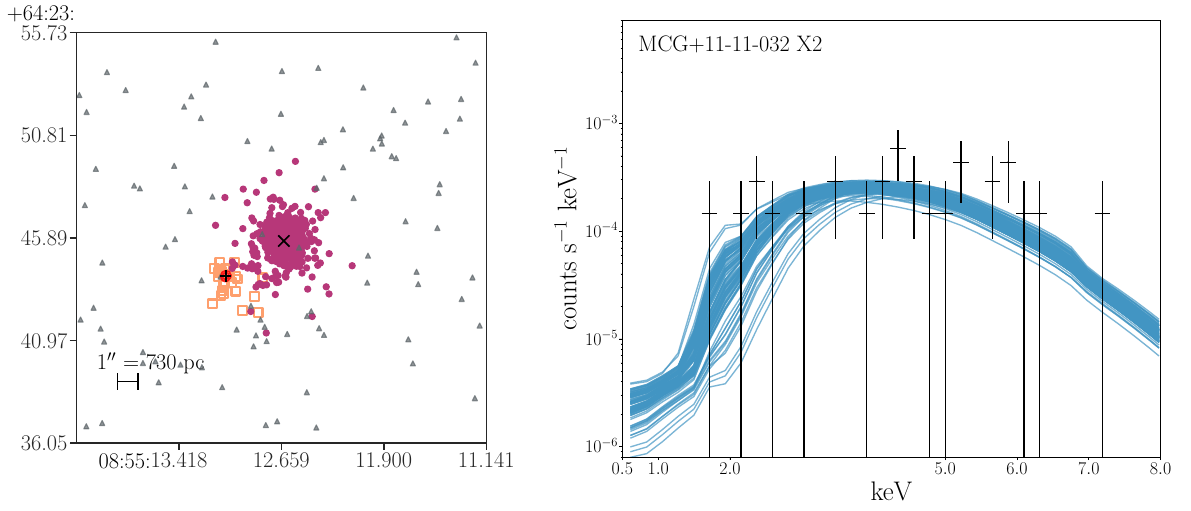}
        \caption{\emph{Left:} The unbinned $0.5$--$8$ keV dataset for MCG+11-11-032, where {\tt BAYMAX} has probabilistically assigns each count to a model component. Here, counts most likely associated with the primary are denoted by red circles, counts most likely associated with the secondary are denoted by open-faced orange squares, and counts most likely associated with background are shown as gray triangles. \emph{Right:} X-ray spectral fits of 100 realizations for the secondary point source (MCG+11-11-032 X2) in MCG+11-11-032. Data have been folded through the instrument response. We overplot one of the spectral realizations in black. The spectra have been rebinned for plotting purposes.}
        \label{fig:countbreakup}
\end{figure*}

We analyze the distributions of the best-fit values for each spectral parameter, and the 2$-$10 keV flux and luminosity.  We define the best-fit value for a given spectral parameter as the median of the distribution of values, and calculate the 99.7\% confidence intervals to estimate the errors. Figure~\ref{fig:countbreakup} shows an example of the X-ray photons probabilisitically assigned to either MCG+11-11-032 X1 or MCG+11-11-032 X2 for one of the realizations.  

In Table~\ref{tab:spectralmodels}, we list the best-fit spectral values returned by our spectral analysis with \BAYMAX. For MCG+11-11-032 X1, we find the distributions for $\Gamma$, $N_{H}$, $F_{\text{2$-$10 keV}}$ , $L_{\text{2$-$10 keV}}$, and parameters associated with both emission lines consistent with the previously calculated best-fit parameters when fitting the extracted spectrum with Model 4. The larger error bars associated with certain parameters from our analysis in Section 2.2 may be due to possible contamination from MCG+11-11-032 X2 when extracting counts within a 2\arcsec\ region centered on the MCG+11-11-032 X1. For MCG+11-11-032 X2, we find the following best-fit spectral parameters: $N_{H} = 9.38_{-2.94}^{+10.30}$ $\times 10^{22}$ cm$^{-2}$; $F_{\text{2$-$10 keV}} = 3.03_{-0.86}^{+0.69}$ $\times 10^{-14}$ erg s$^{-1}$ cm$^{-2}$; unabsorbed $L_{\text{2$-$10 keV}} = 1.52_{-0.48}^{+0.96}$ $\times 10^{41}$ erg s$^{-1}$ (assuming the same redshift as MCG+11-11-032).  Interestingly, the secondary is relatively luminous, nearing the nominal threshold above which classifies persistent X-ray sources as bona fide AGN ($L_{\text{2$-$10 keV}}\sim10^{41} - 10^{42}$ erg s$^{-1}$; \citealt{Fotopoulou2016, Lehmer2019}). We calculate HR=$0.89_{-0.03}^{+0.02}$ and $HR=0.88_{-0.09}^{+0.11}$ for MCG+11-11-032 X1 and X2, respectively.

\subsection{Multi-wavelength Coverage of MCG+11-11-032}
To better understand the possible origin of emission for MCG+11-11-032 X2, we analyze available archival multi-wavelength data of the host galaxy. In the optical regime, there exists SDSS $u, g,r,i,z$ coverage; in the infrared (IR) and ultraviolet (UV) regime there exists Hubble Space Telescope (HST) WFC3/IR F105W ($\lambda_{\text{eff}}=1.04~\mu\text{m}$), WFC3/UVIS F547M ($\lambda_{\text{eff}}=5436~\AA$), and WFC3/UVIS F621M ($\lambda_{\text{eff}}=6209~\AA$) coverage (PropID:12521; PI:Liu); and the radio regime includes the Very Large Array (VLA) Sky Survey (VLASS; 2-4 GHz; \citealt{Perley2011}), the VLA Faint Images of the Radio Sky at Twenty-Centimeters (FIRST; 1.4 GHz; \citealt{Becker1995}), X-band VLA observations (8-12 GHz; ProjectID:13B-020, PI:Comerford), and the LOw-Frequency ARray (LOFAR) Two-metre Sky Survey Data Release 2 (LoTSS DR2; 120-168 MHz; \citealt{Shimwell2022}).

\begin{figure*}
    \centering
    \includegraphics[width=0.9\linewidth]{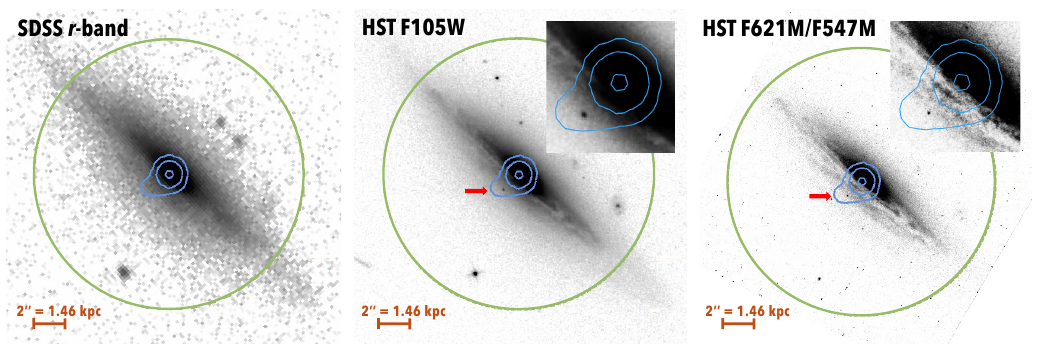}
    \caption{SDSS $r$-band (left), HST WFC3/IR 105W (center), and combined HST WFC3/UVIS F574M/F621M (right) coverage of MCG+11-11-032. Insets for both the center and right panels show a zoomed in field of view of the nucleus of MCG+11-11-032. In green we show the Swifr XRT half-power diameter at 1.5 keV (18\arcsec), which covers the majority of the observable optical light associated with MCG+11-11-032. In blue, we show contours of the 0.5$-$8 keV Chandra emission, with the southeast extension due to the presence of MCG+11-11-032 X2. We find no evidence of an optical counterpart in the SDSS dataset. Within the HST data, all three bands resolve a point source that is coincident with the X-ray coordinates of MCG+11-11-032 (shown with a red arrow). North is up and east is to the left, and a 2\arcsec\ bar (1.46 kpc at $z=0.0362$) is shown to scale.}
    \label{fig:multiwave}
\end{figure*}

We find no evidence of an optical counterpart in the SDSS datasets. However, this is not unexpected given the typical resolution of SDSS ($\sim$1.5\arcsec), and the fact that the projected position of MCG+11-11-032 X2 is coincident with a relatively dense region of the edge-on spiral galaxy, with a highly-obscured line-of-sight. Within the HST data, all three bands resolve a point source that is coincident with the X-ray coordinates of MCG+11-11-032 X2. We note that the level of intrinsic absorption in the spectrum of MCG+11-11-032 X2 is relatively high, on the order of $\sim10^{23}$ cm$^{-2}$. If the object powering MCG+11-11-032 X2 is  also powering the off-nuclear detection in the HST F105W/F621M/F547M data, it is likely that these levels of absorption are circumnuclear. Although the source may sit within a highly obscured line-of-sight, the estimated level of dust attenuation in the V band ($A_{V} = 1.044$ mag; \citealt{Salim2018}) can not account for the level of measured $N_{H}$ alone (e.g., \citealt{Predehl1995}). In Figure~\ref{fig:multiwave} we show SDSS $r$-band, HST F105W, and HST F47M/F621M images.

We find a faint 3.3$\sigma$ noise bump in both VLA survey maps at the coordinates of MCG+11-11-032, however given the low significance of the signal we can not confidently attribute the emission to MCG+11-11-032. Analysis of the X-band VLA observations yield a non-detection at the location of MCG+11-11-032, resulting in an upper-limit on the X-band flux density of approximately 0.13 mJy. Within LoTSS DR2 we find a $\sim 8\sigma$ detection at the coordinates of MCG+11-11-032, with peak flux density of 2.42 mJy/beam. Given the angular resolution of the survey (6\arcsec), the radio detection can be associated with either (or both) of the X-ray point sources. We convert the radio flux density from the LOFAR band to 5 GHz ($S_{\text{5 GHz}}$) by assuming a spectral index of $\alpha=0.7$ (adapting the sign convention $S_{\nu} \propto \nu^{-\alpha}$), which represents the median spectral index value for AGN in LOFAR surveys \citep{CalistroRivera2017}. We estimate a 5 GHz flux density of $S_{\text{5 GHz}} = 0.21$ mJy, consistent with a non-detection in both VLA survey maps ($<0.5$ mJ at 5 GHz) and the X-band upperlimit of $0.13$ mJy at 10 GHz.

\subsection{A Candidate Hyper-luminous X-ray Source}
Given that the X-ray emission of MCG+11-11-032 X2 appears to be off-nuclear, and is relatively luminous, it may originate from an ultra-luminous X-ray source (ULX). ULXs are generally defined as off-nuclear and luminous ($L_{\text{X}} >10^{39}$ erg s$^{-1}$) sources that are distinct from the galaxy's central SMBH (see reviews presented in \citealt{Kaaret2017, King2023}, and references therein). There are various theoretical explanations for the observed emission, the most common being super-Eddington stellar remnants ($M <  10M_{\odot}$). The super-Eddington accretion has been explained via many models, including the presence of strong magnetic fields (e.g., \citealt{Canuto1971, Basko&Sunyaev1976, Bachetti2014, Eksi2015, Mushtukov2015, Tsygankov2016}); geometric beaming \citep{King2001, Poutanen2007, Middleton2015}; and ``leaky disks'' \citep{Begelman2002}. However, the majority of these models predict luminosities $L_{\text{X}} \lesssim 3\times 10^{40}$ erg s$^{-1}$. 

Ever rarer are sources with luminosities above the X-ray luminosity break (XLF) of star-forming galaxies near $\approx10^{41}$ erg s$^{-1}$ \citep{Mineo2012}, classified as hyper-luminous X-ray sources (HLX; \citealt{Matsumoto2003, Gao2003}). For objects with X-ray luminosities above $10^{41}$ erg s$^{-1}$, it is believed that the accretor can be explained by sub-Eddington intermediate-mass black holes (IMBH; \citealt{Taniguchi2000, Colbert&Mushotzky1999, Miller2004, StrohmayerandMushotzky2009, Mezcua2017, Barrows2019, Greene2020, Barrows2024}), where $10^{2} M_{\odot} \lesssim M_{\text{BH}} \lesssim 10^{5} M_{\odot}$. HLXs are rare in the local universe \citep{Gao2003, Sutton2012, Gong2016, Zolotukhin2016}, and searches are hampered by high-levels of contamination from foreground stars and background AGN (\citealt{Zolotukhin2016, Sutton2015}.) 

The strongest candidate for an HLX powered by a massive black hole is ESO 243-49 HLX-1. It has a peak X-ray luminosity of 10$^{42}$ erg s$^{-1}$ and sits approximately 8\arcsec\ (3.7 kpc at $z=0.023$) from the nucleus of the edge-on S0 galaxy ESO 243-49 \citep{Farrell2009}. HLX-1 was shown to be associated with the galaxy using optical spectroscopy \citep{Soria2010} which detected a Balmer H$\alpha$ emission line consistent with the redshift of the host galaxy \citep{Wiersema2010, Soria2013}. Multiwavelength photometric and spectroscopic coverage of ESO 243-49 HLX-1 has allowed for various (albeit, relatively unconstrained) estimates of the black hole mass, ranging from $M_{\text{BH}} \approx 10^{4} M_{\odot}$ \citep{Servillat2011, Davis2011, Godet2012, Straub2014} up to $M_{\text{BH}} < 290 \times 10^{8} M_{\odot}$ (\citealt{Merloni2003, Cseh2015, Webb2018}). It has been proposed that HLX-1 is an IMBH associated with a tidally stripped dwarf satellite galaxy \citep{Webb2010, Farrell2012, Soria2013}, however analyses of the environment around ESO 243-49 have found no evidence of signatures associated with a recent merger event \citep{Musaeva2015, Webb2018}.

If the radio detection in the LoTSS DR2 originates from MCG+11-11-032 X1, we may expect that the mass estimated from the ``fundamental plane of black hole activity'' (FP; \citealt{Merloni2003}) is consistent with the mass of the AGN determined from the CO velocity dispersion of $\log{M_{\text{BH}}/M_{\odot}} = 8.7 \pm 0.3$ \citep{Lamperti2017}. The fundamental plane of black hole activity is an empirical correlation between the mass of a black hole ($M$), the 5 GHz radio continuum luminosity ($L_{\text{R}}$), and the $2–10$ keV X-ray luminosity ($L_{\text{X}}$), reflecting the physical connection between the accretion, outflows, and mass. Offsets from the FP may be expected if the LOFAR detection consists of two radio emitters, or if the LOFAR detection is dominated by radio emission associated with MCG+11-11-032 X2. In the first scenario, we may expect to observe an excess of radio emission with respect to the X-ray luminosity of MCG+11-11-032 X1, resulting in a projected mass that is larger than the previously measured value. In the second scenario, we may expect to observe a lower radio luminosity than necessary to meet the previously measured mass of MCG+11-11-032 X1.

We calculate a 5 GHz radio luminosity of $L_{\text{R}}=3.11\times10^{37}$ erg s$^{-1}$ using our previously derived value for $S_{\text{5 GHz}}$. Pairing this with our estimated X-ray luminosity for MCG+11-11-032 X1, we use the results of \cite{Gultekin2019} to calculate a projected mass for MCG+11-11-032 X1 of $\log{M_{\text{BH}}/M_{\odot}} = 6.27^{+0.71}_{-0.74}$, a value significantly lower than the measured mass. We note that is it unclear if the FP relation is appropriate for systems which are not in a low/hard state, such as Seyferts like MCG+11-11-032 X1. In particular, high accretion rate AGN with radiatively efficient disks likely have quenched their jets, and detected radio emission may be residual from an earlier low/hard state of activity. However, past analyses reflect that the FP may be applicable for high accretion-rate sources (e.g., \citealt{Panessa2007, Gultekin2014, Giroletti&Panessa2009}). With this caveat in mind, we use the FP relation presented in \cite{Gultekin2019} derived using only Seyferts (see their eq. 15). The scatter associated with the FP for Seyferts is larger, likely driven by only having seven Seyferts in their full sample; however, the FP relation for Seyferts versus the full sample remain consistent within $\sim 2\sigma$. We find  $\log{M_{\text{BH}}/M_{\odot}}=5.80^{+1.84}_{-0.59}$, consistent (albeit with larger error bars) with the previous FP mass projection. If we assume that the majority of the radio flux originates from MCG+11-11-032 X2, and use our estimated X-ray luminosity for the source (see Section 3.1), we find $\log{M_{\text{BH}}/M_{\odot}} = 7.29^{+0.45}_{-0.46}$. This value is consistent with the range of masses estimated for HLX-1 using the FP. Previous analyses that resolve the radio emission around ULXs have shown that the radio luminosity can include contributions from large-scale (lobe-like) and a small-scale (jet) structure, resulting in FP projected masses that are biased towards higher values \citep{Mezcua2015}. 

To further investigate the HLX hypothesis, we calculate the ratio of the extrapolated 5 GHz luminosity to X-ray luminosity, $\log{R_{\text{X}}}$ \citep{Terashima&Wilson2003}. This ratio has the advantage over standard optical to X-ray ratios that it can be measured for highly absorbed nuclei ($N_{H} \gtrapprox 10^{23}$ cm$^{2}$). The value of this ratio has been measured to vary for compact objects as a function of their mass, and has been used to flag potential IMBHs (e.g., the ULX IMBH candidate NGC2276-3c, see \citealt{Mezcua2015, Mezcua2013}). In particular, IMBHs may be expected to have values in-between XRBs ($\log{R_{\text{X}}} < -5.3$) and low-luminosity AGN ( $-3.8 < \log{R_{\text{X}}} < -2.8$; \citealt{Mezcua2013}). Once again assuming that most of the radio emission is either associated with MCG+11-11-032 X1 or X2, we calculate $\log{R_{\text{X}}}$ values of -5.4 and -3.7, respectively. MCG+11-11-032 X1 has a value consistent with XRBs -- at odds with its known AGN nature, while MCG+11-11-032 X2 has a value at the lower-end of the AGN range. This may be further proof that the radio emission originates from MCG+11-11-032 X2, and that the source is powered by an IMBH. However, given the unknown relation between the radio and X-ray emission for Seyferts, and thus the unknown expected contribution from MCG+11-11-032 X1 in the LOFAR observation, we can not rule out the possibility that the radio emission is associated with the central AGN.

Perhaps the strongest line of evidence of an HLX origin for MCG+11-11-032 X2 would be the detection of variability in its X-ray luminosity, as the majority of the strongest HLX candidates show evidence of variability on timescales of $\sim$months to a year \citep{KaaretandFeng2007,Lasota2011, Pasham&Strohmayer2013, Wolter2006, Pizzolato2010, Sutton2012}. In particular, HLX-1 has shown to mimic sub-Eddington black hole binaries on the hardness–intensity diagram as it progresses through periodic outbursts \citep{Servillat2011}. Given (1) the low number of counts, (2) the resulting unconstrained X-ray spectrum, and (3) our single epoch Chandra observation, we do not currently have the ability to investigate whether the X-ray emission of MCG+11-11-032 X2 and its variability (or lack thereof) is consistent with other HLX candidates. If MCG+11-11-032 X2 does indeed undergo periodic outbursts, it may be driving the previous claims of a $\sim$25 month periodic behavior in the BAT light curve \citep{Severgnini2018}. However, because the BAT aperture includes both X-ray point sources, it is unclear how their individual behavior would manifest in a joint light curve. Future modeling of the light curve will need to account for two potentially varying X-ray sources.

\subsection{Alternative Hypotheses}
Other origins of emission for MCG+11-11-032 X2 include background AGN. This type of contamination has been proven as the true origin of emission for a previously identified HLX candidate in IC 4320, 2XMM J134404.1-271410. With a measured X-ray luminosity of $L_{\text{X}} \approx 3.5 \times 10^{41}$ erg s$^{-1}$ (assuming the same redshift as IC 4320), it was the second most luminous HLX candidate detected \citep{Sutton2012}. A compact and red counterpart was detected in Very Large Telescope (VLT) VIsible MultiObject Spectrograph (VIMOS) data, and VLT FOcal Reducer/low dispersion Spectrograph 2 (FORS2) long-slit spectroscopy proved the source was a background QSO at $z=2.84$ \citep{Sutton2015}. Using the $\log{\text{N}}-\log{\text{S}}$ cumulative number density of cosmic X-ray background sources from deep field surveys \citep{Bauer2004}, we quantify the probability that MCG+11-11-032 X2 is a background source. We find the  probability of background contamination within a 3.3\arcsec\ radius around the center of MCG+11-11-032 is negligible ($P=8\times10^{-3}\%$). We note that a similar type of analysis carried out for 2XMM J134404.1-271410, which sits 18\arcsec\ from the center of IC 4320 yield a significantly higher probability ($P\sim 0.3\%$). However, analysis of the optical spectrum is necessary to confidently rule out contamination from a background QSO.
Lastly, we note that the extended X-ray emission is likely not associated with a jet, given the coincident point-like sources detected in Chandra and HST. No other evidence for jet morphology is resolved in the available SDSS and HST images. However, higher-resolution radio coverage can provide better insight on whether the properties of MCG+11-11-032 X2 are consistent with a jet.

\section{Conclusions}
In this study, we present an X-ray analysis of MCG+11-11-032. Using new Chandra observations, we investigate two hypotheses for both a binary (at sub-pc separations) and a dual (separated by 0.77\arcsec) AGN in the nucleus of MCG+11-11-032. Chandra observations uniquely allow for resolving any two AGN in a dual system and/or confirming the presence of any two narrow Fe emission lines at $\sim6.4$ keV in a binary system. Additionally, new Chandra observations allow for further unexepcted discoveries. Multi-wavelength data provided by SDSS, HST, VLA, and LOFAR enable a comprehensive analysis of any X-ray detections. The main results of this study are summarized below: \\
\begin{enumerate}
\item Analysis of the dataset with \BAYMAX yields a Bayes Factor $\logBF = 37.87 \pm 1.95$ in favor of the dual point source model. The position of the secondary X-ray point source -- MCG+11-11-032 X2 -- coincides with the position of the visibly extended emission, and the best-fit separation and count ratio returned by \BAYMAX are $r=3.27\arcsec_{-0.27}^{+0.32}$ (approximately 2.7 kpc at $z=0.0362$) and $\log{f}=-1.86_{-0.28}^{+0.28}$. The calculated position angle between the two point sources is $\text{PA}_{\text{X}}=149\pm{4}^{\circ}$, which is approximately perpendicular to the position angle of the galaxy's plane. Given that the best-fit coordinates of the two X-ray sources correspond to a separation over 4 times as large, and a position angle that is perpendicular, to the values previously reported in \cite{Comerford2012}, it is unlikely that this source is the origin for the observed double-peaked [\ion{O}{3}] emission lines.
\item We find no evidence of double-peaked narrow Fe K$\alpha$ lines with an orbital velocity offset of $\Delta v \approx 0.06$ in the X-ray spectrum of MCG+11-11-032. The X-ray spectrum is best fit by an absorbed powerlaw with two emission lines: a single Fe K$\alpha$ line centered at $6.33_{-0.12}^{+0.21}$ keV, and an emission line centered at $7.56_{-0.31}^{+0.18}$ keV, which can possibly be associated with either Fe K$\beta$, Ni K$\alpha$, or a combination of the two. We conclude that the Chandra data do not support the binary AGN interpretation.
\item Further investigation of the Chandra observation shows evidence for soft extended X-ray emission at energies $<$2 keV and angular scales $\approx$4\arcsec. However, given the low number of counts associated with this component ($\sim 12$) it is difficult to accurately constrain the geometry and spectrum. We test the legitimacy of the assumption that MCG+11-11-032 X2 is a hard compact point source, separate from the soft extended X-ray emission via two analyses: re-running \BAYMAX on only the hardest photons (3$-$8 keV) and re-running \BAYMAX using updated models that account for an additional high-count diffuse emission component. Both tests yield strong evidence in favor of the dual point source model.
\item We use the results from \BAYMAX to carry out a spatially resolved and statistical spectral analysis of both the central AGN (MCG+11-11-032 X1) and the secondary X-ray point source (MCG+11-11-032 X2). On average, the spectrum MCG+11-11-032 X2 only has 27 counts between 0.5$-$8 keV, while MCG+11-11-032 X1 has 1965. For MCG+11-11-032 X2, we find the following best-fit spectral parameters: $N_{H} = 9.38_{-2.94}^{+10.30}$ $\times 10^{22}$ cm$^{-2}$; $F_{\text{2$-$10 keV}} = 3.03_{-0.86}^{+0.69}$ $\times 10^{-14}$ erg s$^{-1}$ cm$^{-2}$; unabsorbed $L_{\text{2$-$10 keV}} = 1.52_{-0.48}^{+0.96}$ $\times 10^{41}$ erg s$^{-1}$ (assuming the same redshift as MCG+11-11-032 X1). Interestingly, the secondary is relatively luminous, nearing the nominal threshold above which classifies persistent X-ray sources as bona fide AGN ($L_{\text{2$-$10 keV}}\sim10^{41} - 10^{42}$ erg s$^{-1}$).
\item Hubble Space Telescope imaging in IR F105W ($\lambda_{\text{eff}}=1.04 \mu\text{m}$), and UV F621M/F547M ($\lambda_{\text{eff}}=6209 \AA$ and $\lambda_{\text{eff}}=5436 \AA$, respectively) bands resolve a point source that is coincident with the X-ray coordinates of MCG+11-11-032 X2. If MCG+11-11-032 X2 powers the emission detected in both HST and Chandra observations, it may be a hyper-luminous X-ray source. 

\item We find an $8\sigma$ LOFAR detection from the LoTSS DR2 survey at the coordinate of MCG+11-11-032. Assuming the radio emission is dominated by the central AGN, the fundamental plane of black hole accretion predicts a projected mass that is inconsistent with, and significantly lower than, the previously constrained mass of the central AGN. Additionally, we calculate a radio to X-ray ratio $\log{R_{\text{X}}}=-5.4$, consistent with an XRB and at odds with its known AGN nature. However, if the radio emission is dominated by MCG+11-11-032 X2, the fundamental plane of black hole accretion estimates a projected mass of $\log{M/M_{\odot}} = 6.27_{-0.74}^{+0.71}$, consistent with the estimated mass for the best IMBH/HLX candidate to date, HLX-1. Additionally, we calculate a radio to X-ray ratio of $\log{R_{\text{X}}} = -3.7$, a value consistent with the lower-end of the AGN range. We note that a spatially resolved optical spectrum of the source is necessary to confidently rule out contamination from a background QSO.
\end{enumerate}

Further data that could support the hypothesis that emission of MCG+11-11-032 X2 is associated with an HLX include (1) a spatially resolved optical spectrum of MCG+11-11-032 X2 that confirms its association with the galaxy MCG+11-11-032; (2) high-resolution and deep radio observations that can spatially resolve the radio emission around MCG+11-11-032 X1 and X2; and (3) the detection of X-ray variability on timescales of months associated with varying spectral hardness. Regarding (1) and (2), planned follow-up observations with ground-based, such as the VLA and Integral Field Unit spectrographs such as the Multi-Unit Spectroscopic Explorer (MUSE), will better constrain the origin of the X-ray emission associated with  MCG+11-11-032 X2. Regarding (3), the Chandra X-ray observatory is the sole telescope that can resolve sources separated on scales $<6$\arcsec\ and generate an individual lightcurve for MCG+11-11-032 X2. We emphasize that  the detection of MCG+11-11-032 X2 was only possible due to Chandra's superb angular resolution, highlighting how previous binary AGN interpretations from larger-aperture instruments, such as Swift, need careful evaluation and further follow-up. This is one of many examples where high spatial resolution in the X-ray band is extremely important; wide field-of-view but lower angular resolution X-ray imagers are not best-suited to complement the current and near-future fleet of NASA's observatories with sub-arcsecond spatial resolution such as the James Webb Space Telescope and the Nancy Grace Roman Space Telescope. It may not be until the early- to mid-2030s when we will have access to another high-resolution X-ray imager, such as AXIS \citep{Reynolds2023} or HEX-P \citep{Morrissey2023}, that will be able to resolve closely-separated compact objects.

\begin{acknowledgments}
M.M. acknowledges support from the Spanish Ministry of Science and Innovation through the project PID2021-124243NB-C22. This work was partially supported by the program Unidad de Excelencia Mar\'ia de Maeztu CEX2020-001058-M. F.M-S. acknowledges support from NASA through ADAP award 80NSSC19K1096. The National Radio Astronomy Observatory is a facility of the National Science Foundation operated under cooperative agreement by Associated Universities, Inc. Basic research in radio astronomy at the U.S. Naval Research Laboratory is supported by 6.1 Base Funding.
\end{acknowledgments}
%

\vspace{5mm}
\section*{Data Availability}
All the {\it HST} data used in this paper can be found in MAST: \dataset[10.17909/3xyk-1f32]{http://dx.doi.org/110.17909/3xyk-1f32}
\facilities{CXO, SDSS, HST, VLA, LOFAR}


\software{{\tt CIAO} (v4.12; \citealt{Fruscione2006}), \newline XSPEC (v12.13.1; \citealt{Arnaud1996}), \newline {\tt nestle}~(https://github.com/kbarbary/nestle), \newline {\tt PyMC} \citep{Salvatier2016}, \newline {\tt MARX} (v5.3.3; \citealt{Davis2012})}


\appendix 
\begin{figure*}[h]
    \centering
    \includegraphics[width=0.9\linewidth]{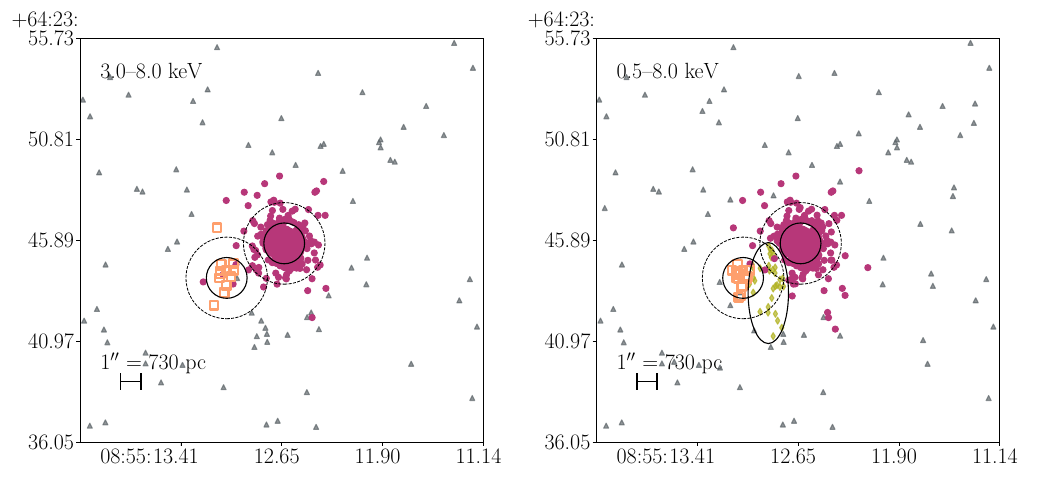}
    \caption{The unbinned 3$-$8 keV (left) and 0.5$-$8 keV (right) dataset for MCG+11-11-032, where we use \BAYMAX to probabilistically assign each count to a different model components. Here, counts most likely associated with the primary are denoted by red circles, counts most likely associated with the secondary are denoted by open-faced squares, counts most likely associated with background are denoted as gray triangles; in the right panel, counts most likely associated with a high-count diffuse emission component are shown in green diamonds. In solid (dashed) black lines we plot 1\arcsec\ (2\arcsec) radius circles centered on the best-fit location for the primary and secondary point sources. When analyzing the 3$-$8 keV dataset with \BAYMAX, which excludes photons associated with an extended and soft component, we find a $\log{BF}=9.38 \pm1.95$ in favor of the dual point source model. The position of the secondary is consistent with our findings when analyzing the full data-set (see Section 3.1 for more details). When analyzing the full 0.5$-$8 keV dataset with an additional model component that accounts for a high-count diffuse emission region (sitting within the black ellipse shown on the right panel), we find a Bayes factors of $\log{BF} = 28.6 \pm 2.2$ in favor of the dual point source model. Even with an additional high-count component included in our models, we find strong evidence for the presence of a point source.}
    \label{fig:appendix}
\end{figure*}

\bibliography{main}{}
\bibliographystyle{aasjournal}



\end{document}